# The Future is Meta

*Metadata, Formats and Perspectives
towards Interactive and Personalized AV Content*

Alexander Weller,[1] Werner Bleisteiner,[2] Christian Hufnagel[3,] Michael Iber[4]

*[1] Südwestrundfunk, Fachhochschule St. Pölten, alexander.weller@swr.de*
*[2] Bayerischer Rundfunk, werner.bleisteiner@br.de*
*[3] Südwestrundfunk, christian.hufnagel@swr.de*
*[4] Fachhochschule St. Pölten, michael.iber@fhstp.ac.at*

**Abstract**

The production of media content has undergone tremendous changes in recent years. Multiple daily content updates are just as common for some platforms as is processing the provided content specifically for their target audiences. Such features are made possible through metadata, which make information accessible by categorizing it. In conjunction with AI-supported tools, metadata are shaping the future of audio-visual content production, distribution and consumption. It allows editors to effectively search through archives like in the Tailored Media Project, broadcasters to provide content that is adapted to users' surroundings like in the ARD Audiothek unterwegs project, or give users the ability to experience audio-visual content from different perspectives like in the ORPHEUS project. Although these projects provide comprehensive insight into the potential of metadata, their integration in existing infrastructures meets several limitations. For example, content-related metadata may initially be generated at some point during the production process but will then be lost at later stages due to current standards and incomplete software implementations. In our contribution, we will discuss requirements and potential approaches and give an outlook on possible fields of application and use-cases.

## 1. Introduction

The enormous influence of Artificial Intelligence on our entire culture, particularly affects the field of media. Impacts relate to all aspects of the media product life cycle, starting from production, processing, distribution, and archiving to the use of the final product by the consumer. The manifestations of today's products frequently differ significantly from their historical predecessors. This can be seen in the example of a pop album on a vinyl record which can be understood as a self-contained entity of songs to be played in a predetermined order. The record is only available as a whole and skipping a song is only possible with rather cumbersome manual intervention. With the introduction of the CD player and its shuffle playback function, at least the order of the tracks can be changed randomly. Today's streaming services have changed the described situation fundamentally. The entity of the record has given way to the complete catalog of the music platform. This has gone along with a fundamental change of listening habits. Rather than listening to complete albums, consumers tend to listen to often even automatically generated playlists containing a potpourri of songs by different artists. In order to meet their customers' musical tastes and preferences, music providers implemented algorithms to make suitable recommendations based on individual listening habits. On the one hand this requires collecting information about the customers' preferences, on the other hand it is also required to have information about the music itself, which can include descriptions, genres, tempos, etc. In both cases this information is called metadata.

In this article, we approach the topic of metadata from the perspective of audiovisual production and describe on the basis of a use case
- how metadata is implemented in the workflows of public-service broadcasting,
- what challenges still need to be addressed, and
- what perspectives are opening up for the future.

In the following chapter, we will provide a general definition of metadata. Chapter 3 focuses on metadata formats in audiovisual (AV) broadcasting. In Chapter 4, we refer to some application examples that have been implemented at a prototype stage. This is followed by a closer examination of the current challenges within the complex environment of a large-scale infrastructure, such as ARD and its nine self-governing regional broadcasters. Chapter 5 provides an overview of new approaches that offer greater flexibility for embedding metadata, as well as AI-based tools that streamline the workflow and assist in generating a wide variety of metadata automatically. We will also offer a glimpse into the future of metadata usage within broadcast production.

## 2. Data about Data

Although ubiquitous nowadays, the concepts of metadata still carry an aura of the intangible, the indeterminate, even the mysterious. This may be partly due to the fact that metadata are primarily mentioned today in connection with the equally enigmatic field of artificial intelligence, whose complex algorithmic interconnections and network layers even experts can no longer fully comprehend. Yet, metadata them-selves are something rather mundane that have been firmly established in our socio-cultural environments for centuries. They are essential elements of everyday life without which we would hardly be able to locate a book in a library by knowing only its author and title, or find out whether it is available or borrowed by someone else.

Over time, the literal translation of the ancient Greek prefix 'meta,' meaning 'beyond,' has been increasingly associated with the concept of 'about.' Therefore, metadata can be understood as 'data about data,' a definition that may demystify the term to some extent.

## 2.1 Types of Metadata

### Descriptive Metadata

As previously illustrated using the example of a book in a library, several categories of metadata exist. According to Gartner [1], data about the book – including title, author, publication date, ISBN, and keywords, which pertain to topic and content-related information – fall under descriptive metadata. Whether we have precise information about the book we're seeking or just a vague notion of its content, metadata should facilitate our search for exactly what interests us and also allow us to exclude what we're not looking for.

### Administrative Metadata

On the other hand, administrative metadata ensure that we know the book's location in the library. They inform us whether it's currently on loan and to whom. Technical metadata are a subcategory of administrative metadata, providing details about the format and compression of stored audio files to ensure proper playback on media players. Rights metadata also fall under administrative metadata, serving to store information regarding intellectual property rights and access rights for users, possibly in connection with billing models (for instance, 'pay per view').

### Structural Metadata

In addition to descriptive and administrative metadata, a third primary category is structural metadata. In the form of page numbers, this type of metadata ensure that we can read a disordered manuscript in the correct order, or they are used to synchronize a film with its soundtrack through timecodes.

## 2.2 Linking Data to Wisdom

For Gartner [1], the significance of metadata extends far beyond mere object or information retrieval:
> Metadata has a role that is more important than simply allowing a single fragment of information to be stored away and found again. It enables these to be linked together to form knowledge and for this knowledge to be consolidated into what we understand as culture. [1, p. 9]

Culture is also the essence of Ackhoff's Pyramid (cf. Fig. 1) [2 quoted according to [1]] to which Gartner refers when he describes metadata as connecting links of the five levels of the pyramid: data, information, knowledge, understanding, wisdom.

Returning to our example of a music recommendation system from Chapter 1, we will take a closer look at how metadata can be used to turn raw data into wisdom. This follows the verified and approved output of ChatGPT-3.5 on the prompt: 'How can the Ackoff pyramid (data, information, knowledge, understanding, wisdom) be used to explain a music recommender system, using the term metadata to depict the transitions between the five layers?':

> The Ackoff pyramid (data, information, knowledge, understanding, wisdom) can be applied to explain a music recommender system, with metadata illustrating the transitions between the layers as follows:
> **Data:** At the base of the pyramid, we have raw data. In the context of a music recommender system, this represents the vast collection of songs, their attributes (metadata), and user interactions with the system, such as song ratings, plays, and skips.
> **Information:** Metadata is used to structure and organize the raw data, turning it into meaningful information. This metadata includes song attributes like genre, tempo, artist, album, and user data, such as preferences, listening history, and demographic information. The system uses this metadata to create a structured database of music content.
> **Knowledge:** As the system processes the metadata, it recognizes patterns and relationships within the data. For instance, it may identify that a user often listens to rock and pop songs or that certain songs share similar attributes. This knowledge allows the system to make informed music recommendations based on user preferences and metadata-derived insights.
> **Understanding:** Understanding emerges when the system comprehends the deeper context behind the knowledge gained from the metadata. This involves recognizing that a user's musical taste is not limited to a single genre and that their preferences evolve. The system understands user context and tailors recommendations accordingly, leveraging metadata insights to provide a more personalized experience.
> **Wisdom:** Wisdom is the highest level, where the system not only understands the user's current preferences but can also adapt and make decisions that align with the user's evolving tastes and moods. The system leverages its deep understanding of user preferences, combined with metadata, to offer recommendations that are not only in line with the user's known preferences but can also introduce new and relevant music to enhance the user's music experience.

In this context, metadata serves as the key transition point between data and information, playing a crucial role in organizing, categorizing, and enriching the raw data. It's the metadata that allows the music recommendation system to progress through the Ackoff pyramid, ultimately providing users with personalized and meaningful music recommendations based on knowledge, understanding, and wisdom derived from the metadata-rich dataset.

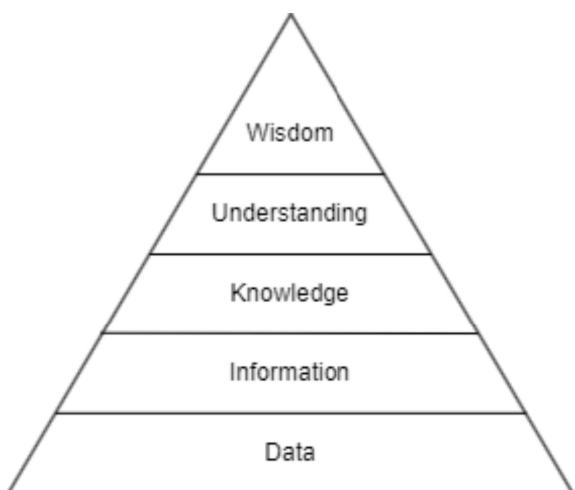

Fig. 1: The pyramid represents the relationships between data, information, knowledge, understanding, and wisdom as described by Ackoff [2].

## 2.3 Manual and Automatized Metadata Generation

For universal usage of metadata, they should adhere to binding standards. To retrieve the attributed objects or described data, it is crucial that they follow a clear set of rules in terms of their syntax and semantics.

Metadata can be created manually through text entry or automatically using algorithms. Manually generated metadata carry the risk of including subjective viewpoints in the description, as well as typos or syntax inconsistencies that can hinder later retrieval. One challenge of algorithmically generated metadata is ensuring the accurate recognition and attribution of analyzed data, a potential source of error that persists even with the increased use of advanced machine learning and complex neural networks.

## 3. Metadata in Audiovisual Broadcasting

Since the 1990s, as broadcasting underwent a fundamental digital transformation [3], the necessity to collect and structure audiovisual (AV) metadata has led to the development of various methodologies and classifications. One significant outcome of this process was the emergence of 'ontologies,' which are formal, explicit specifications of shared conceptualizations [4].

According to Zeng and Qin [5], an ontology describes:
- the types of things that exist (classes)
- the relationships between them (properties)
- the logical ways those classes and properties can be used together (axioms)

Throughout the last few decades, numerous ontologies have been created for diverse purposes and ecosystems, with earlier ontologies primarily concentrating on archiving processes [5].

Three notable metadata ontologies within the broadcasting domain are: MPEG-7, EBU-Core and BMF.

### MPEG-7

MPEG-7 was created to describe multimedia content such as digital images, digital video or digital audio in a complete way [6]. It has been standardized as ISO/IEC 15938 [7].

### EBU-Core

EBU-Core is a metadata schema based on the Dublin Core (DC) with the intent to to create a set of descriptive, administrative, technical and structural metadata for audiovisual content [6]. The goal was to create metadata descriptions for a wide range of broadcasting applications such as archives, exchange and production. It has been specified in EBU Tech 3293 [8].

### BMF

The Broadcast Metadata Exchange Format (BMF), was developed by IRT for german-speaking broadcasters such as ORF, ARD and SRG SSR [9]. It was developed as a framework for metadata exchange in broadcast environments. It contains descriptions for television, radio and online production.

## 3.1 Types of Audiovisual Metadata

The terminology and definition of metadata in the area of professional AV production slightly differ from the general descriptions given in Chapter 2.1. According to Dickreiter et al. [10], three different categories of metadata can be identified for describing AV content. These categories can either be embedded, along with the so-called 'essence' (i.e., the content described by the metadata), in the media file, or they can be stored externally in additional files (cf. 3.2.2).

**Technical Metadata**

For technical metadata, the description corresponds to the one given in Chapter 2.1. Technical metadata describe the technical parameters of the essence, such as codec, bitrate, resolution, or sample rate. Since this information is required, for instance, for the playback and further processing of audio files, technical metadata are always embedded within the file itself and/or the encapsulating media container format (cf. 3.2.1), which merges several audio and video streams together.

**Formal Metadata**

Formal metadata contain factual information about media content, such as title, author, composer, artist, or producer. These are indispensable for content identification, rights management, and distribution. A minimum set is, at best, at least partly found in the file, for example, within its name. In itself, and according to the definitions given in Chapter 2.1, formal metadata would be considered subcategories of both descriptive and administrative metadata. However, in AV production, as shown in the following paragraph, descriptive metadata rather represent qualitative content depictions, which may justify the deviation from general metadata terminology.

The standardization of file formats does not necessarily include the complete implementation of formal metadata methodologies. Partial or incomplete integration of metadata structures often prevent a smooth exchange between different systems and impede both the production process and comprehensive innovative approaches. As the necessary data-set within the overall business process might be quite extensive, the amount of metadata might be in conflict with file size limitations of the used format (cf. 3.2.1). To prevent this issue, the complete set of formal metadata is transferred to a Media Asset Management System (MAM) or Content Management System (CMS)(cf. 3.2.2). A MAM system is used to manage rich media assets throughout a company spanning multiple business processes. A rich media asset could be for instance, a media asset such as multimedia content enriched with its rights information [11]. In comparison, a CMS is used to manage assets used for websites, such as templates, images or videos, for example for interactive websites where specific information is stored in a database and dynamically loaded into the site upon access [11].

**Descriptive Metadata**

Descriptive metadata refer to interpretative attributes of media content, such as type, category, genre, or mood. In their denotation, they adhere to the general categorization of metadata presented in Chapter 2.1. Unlike formal metadata, descriptive metadata in the context of AV media are subject to at times subjective interpretations of attributes, including re-evaluation, alteration, and re-interpretation procedures, depending on the targets of curation and distribution. For example, the keywords for a song might differ if the target audience is the general public as opposed to musicologists. Since descriptive metadata tend to become rather extensive, they are commonly stored separately within additional files in databases to avoid conflicts with file size limitations and restrictions regarding the specifications that may apply to file formats using embedded metadata (cf. 3.2.1).

## 3.2 AV Media Formats and Metadata Storage

### 3.2.1 Internal Metadata Storage

As mentioned in the previous subsection, the design of AV media file formats and containers, as used in professional production environments, allows for the embedding of additional information alongside the essence. The metadata can be stored either in the file itself or in a container format. A container format enables the combination of multiple data streams along with corresponding metadata into a single file, such as MXF in a professional setting or MP4 in a consumer setting. However, consumer formats are not addressed in this publication.

The most common AV media file formats used within the network of the European Broadcasting Union are:

**BWF & BW64 (Audio-Only)**

The Broadcast Wave Format (BWF) is an extension of the RIFF-WAV format, initially developed to store uncompressed audio. BWF was originally standardized in 1997 by EBU Tech 3285 [12] and subsequently as ITU-R BS. 1352 [13]. One of its major restrictions is the limitation of the file size to 4 Gigabytes. Additionally, BWF is limited in terms of advanced multichannel audio and its inadequate support for technical metadata. A BWF file can contain a description of the content embedded in the <bext> chunk, but it is limited to 256 ASCII characters.

In order to overcome these file size restrictions RF64 was established in 2005, as specified in EBU Tech 3306 [14]. As a 64bit file format, RF64 can address 16 exabyte of data, which equals 16 Billion gigabyte [15]. This translates to over 2 million years of stereo audio at 24 bit and 48 kHz. To solve the metadata restrictions present in BWF, BW64 was developed. It features a 64-bit address and supports the embedding of metadata based on the Audio Definition Model (ADM), as well as any XML-formatted metadata in its <axml> 'chunk', i.e., a specific data block within the file [16].

ADM enables the embedding of metadata that describes advanced multichannel audio. In combination, BW64 and ADM support channel-based audio, as well as scene-based audio, and object-based audio, all within the same file. Channel-based audio is used as an audio representation to describe elements where each track represents a channel (e.g. Stereo, 5.1, …), scene-based audio is used to describe elements where each track corresponds to coefficient signals such as ambisonics, and object-based audio is used to describe an audio representation in which the elements of the content (e.g. Dialogue, Music, Ambience, …) are separate, but accompanied by metadata describing their relationship [17].

This setup is accompanied by metadata that describes each part of the audio, i.e., each track is detailed individually and more comprehensively compared to BWF. A BW64 file has no restrictions on the size of the <axml> chunk, meaning that the description of the content is only limited by the file size restrictions of BW64, which is – just as RF64 – 16 exabytes.

Problems of backwards compatibility and transferability of metadata between BWF and BW64 have been addressed and solutions are currently in the process of being implemented.

### MXF (Audiovisual)

The Material Exchange Format (MXF) is a video-centered container format that has been specified in multiple versions by the Society of Motion Picture and Television Engineers (SMPTE), covering various requirements and applications, such as broadcast specific file and metadata requirements [18].

MXF enables the encapsulation or 'wrapping' of multiple audio and video essences together based on application-specific requirements. These so called 'Operational Patterns' allow for a application specific array of data and metadata to be contained in an MXF file. Within Europe, the most commonly used version for broadcast video exchange is MXF OP1A: SMPTE ST 378 [19].

### 3.2.2 External Storage

While technical metadata are always integral parts of the media file, formal and descriptive metadata are not necessarily embedded. Initially, professional AV media file formats focused on the technical quality of the essence, the media content itself, while little attention has been dedicated to the challenges of accommodating extensive formal and descriptive metadata within the files themselves. Therefore, embedding comprehensive metadata directly within the files has not been widely considered. Instead, the use of unique identifiers has become necessary to link essences with their metadata stored in separate files. This also requires additional steps to ensure that both files are forwarded synchronously within complex networked infrastructures. To enable the storage and editing of essences alongside their metadata, developments of Media Asset Management (MAM) systems and Content Management Systems (CMS), as well as connected database systems, were necessary.

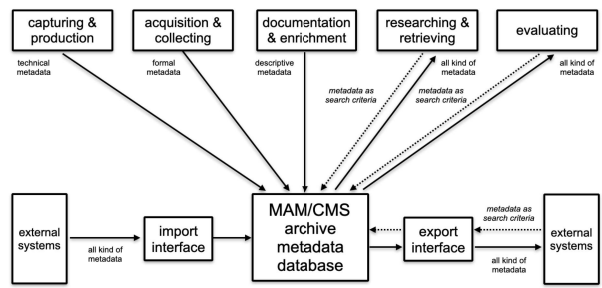

Fig. 2. A generalized representation of a metadata workflow describing how the involved systems are connected, what type of metadata they transport, and where they are stored [10].

## 4. Metadata Guided Audio (MGA)

The integration of metadata into existing and future formats, workflows, and production networks is one of the key endeavors of Metadata Guided Audio (MGA)[20]. The term MGA typically describes emerging audio experiences that do not (solely) rely on channel based audio, such as stereo or 5.1 [21]. Instead, MGA aims to provide listeners with acoustic experiences that are more accessible, personalized, and interactive, regardless of the consumption method, be it through headphones, soundbars, or multi-speaker setups [22]. Previously, the term Next Generatio Audio (NGA) was used to describe this endeavor, but the definition has since changed to MGA describing audio essences accompanied by descriptive metadata e.g. ADM and S-ADM [20]. In comparison, NGA is now used as a generic term to describe specific delivery technologies and codecs that can combine object-based, scene-based and channel-based audio e.g. MPEG-H and AC-4. The term MGA was first used by SMPTE during the process that lead to the standardization of ST2127-1 [23] and ST2127-10 [24] that describe the mapping of ADM metadata into MXF.

MGA extends far beyond purely technical solutions and paves the way for new content formats and play forms. In the following paragraphs, we will present some of these new concepts, which up to now have only been realized as prototypes.

### 4.1 Paradigmatic Pilot Projects of MGA

Over the past decade, several paradigmatic pilot projects have been conducted, highlighting the potential for creating immersive, adaptive, and interactive sound experiences by utilizing metadata.

In 2012, the Research & Development department of the BBC (BBC R&D) initiated several experiments to innovate established content formats by integrating the technological advancements available at the time.

One of the experimental outcomes was the publication of an adaptive version of the radio documentary "Derek Tangye - The Cornish Gardener" on BBC TASTER [25], the BBC's experimental platform. Originally produced by John McCarthy for BBC Radio 4, this documentary explored the life of British author Derek Tangye, known for his series of books, 'The Minack Chronicles' [26].

By 'reverse engineering' the audio content, the documentary was analyzed and divided into segments, making its length and informational depth variable [27]. This variability, a core concept of Metadata Guided Audio (MGA), enables users to adjust the content to fit specific time spans — for example, the duration of a train journey — or to tailor it to their preferred depth of information, as seen in a personalized news reel.

From 2015 to 2018 the EU funded research project ORPHEUS [28] further explored the possibilities of MGA. The consortium consisted of ten companies (Fraunhofer Gesellschaft zu Förderung der angewandten Forschung, European Institute for research and strategic studies in telecommunications (Eurescom), British Broadcasting Corporation (BBC), Institut für Rundfunktechnik (IRT), Guppies in the dark, Trinnov Audio SA, b-com, Institut de recherche et de coordination acoustique musique (IRCAM), Bayerischer Rundfunk (BR), and Magix Software) to pioneer the development of an end-to-end object-based audio chain.

One of the subprojects aimed to incorporate features of variable length and information depth into a prototype app [28]. However, as the project's main goal was to develop an end-to-end object-based audio chain, an elaborate approach like that applied in 'The Cornish Gardener' was out of scope. Instead, the principle of 'Levels of Importance (LOI)' was applied to the semantic segments of an audio piece. LOI is a principle derived from information transmission and journalistic practice, known as 'The Inverted Journalistic Pyramid.' [29]. In this model, the most important information in a news segment is placed at the beginning of an article, at the top of the pyramid (the so-called 'lead,' providing key information: 'who,' 'what,' 'when,' 'where,' 'why,' and 'how'), followed by crucial information that is slightly less important (the 'body,' adding context and background). At the bottom of the pyramid, the least important information is found (the tail, e.g., related details)[30]. Based on this information stored as metadata alongside a news program, the content delivered to the user can be adapted in length by removing segments with lower LOI or adding segments with higher LOI.

The ORPHEUS variable length approach has been further developed by one of the authors, who created a plug-in for segmentation in the Digital Work Station (DAW) 'Reaper' [31], developed a metadata structure, and designed a compatible audio player for the web browser. [32]. The approach is particularly suitable for podcast production, allowing editors to create and test variable-length content, and enabling users to listen to variable-length audio content on-demand. The metadata necessary for the variable-length playback is generated based on the timecodes and labels of the markers that the editors set in their DAW.

In 2019, the various attempts and pilots projects for object-based media developed by BBC R&D culminated in the MAKER BOX [33], which serves as a central hub for the creative community to 'discover, learn, and connect.' Tools such as 'StoryFormer', 'Audio Orchestrator', and 'Adaptive Podcasting', which are currently available, enable new ways of experiencing audio and heavily rely on metadata. Audio Orchestrator, for instance, allows users to utilize their smart devices to create individually tailored spatial audio environments. To function effectively, a wealth of metadata is required to direct the correct audio to the appropriate smart devices, determining factors such as which audio plays from a specific direction, which device should play this audio, and how the audio needs to be adapted to fit the device and its position. StoryFormer is a tool designed to develop flexible, responsive videos. Based on viewers' input, the content can be dynamically adapted to their preferences, context, and choices. A key point for this product's effectiveness is the foundation of metadata, which pertains to the available choices, where those choices lead (i.e., to which files), and which files serve as alternatives.

Since 2022, the MAKERS BOX has also included an 'Adaptive Podcast Player' with an adaptable editing platform, allowing content creators to publish podcasts that adapt to phone sensor inputs, such as the time, location, or brightness, rather than offering a linear experience. Most recently, the source code of this project has also been published on GitHub [34]. To enable adaptive playback based on sensor input, metadata about the podcast need to be accessible. This includes not only traditional podcast metadata, such as file locations, but also metadata about how the adaptive podcast is constructed. It should detail which sensors it relies on at various points in time and how it responds to these inputs.

In 2021, the broadcaster BR developed 'Drive by Bayern,' now called 'ARD Audiothek unterwegs', a prototype that allows personalized audio experiences in a user's car. By accessing the location metadata stored alongside each news segment in their archive, they can offer users 515 content suggestions based on, for example, historical places visible from their car while driving [35].

Another prototype related to MGA is the 'Regional Update,' an approach by BR from 2023, where users can listen to a news mix personalized to their location [36]. They can select a town, specify a time range for how recent the news should be, and set a limit for how far away the news can originate. Based on these parameters, a personalized mix of news is created. To facilitate this prototype, location metadata had to be AI-generated and stored for each news segment produced by every local broadcaster.

TailoredMedia, a collaboration between Joanneum Research, ORF, the Austrian Mediatheken, and FH St. Pölten, specifically addresses journalists and archivists [37]. In this project, methods for the automatic analysis of audiovisual content using artificial intelligence (AI) were developed and explored. This automation facilitates the enrichment of content, including video, with semantic metadata e.g. for media monitoring, journalism, and archiving.

## 4.2 Implementation of Personalization Concepts in Existing Production Workflows

While the development of innovative approaches is quite advanced, implementing metadata structures in existing production processes and systems presents particular challenges. The available options of individual systems may be outdated in this regard, or existing standards may only be partially integrated. Furthermore, the bandwidth of existing interfaces between systems might be limited, allowing only a restricted flow of information. In addition to these challenges, consider the case of ARD Mediathek (the central video-on-demand platform of ARD) and ARD-Klassik (a genre-specific subsection of ARD Mediathek), which involve several regional public broadcasters (LRAs) of the federal network, each with different backgrounds and infrastructures. Given these complexities, it is understandable that the task of homogenizing the production workflow is far from trivial.

In the following, we will present a case study on the production workflow and metadata processes at ARD-Klassik, as an integral part of the ARD Mediathek on-demand hub [38]. This case study focuses on the current infrastructure, processes, and workflows involved in recording a classical concert within various branches of the federal broadcasting network. It begins with the editorial departments of the LRAs and extends to the storage of these recordings in the ARD Mediathek. These insights will form the basis for analyzing the current challenges that arise in the context of these processes and the network infrastructures. To limit the scope of this rather complex undertaking, our focus is restricted to on-demand content. Additionally, workflows that facilitate the recordings or live-streaming processes will not be discussed.

ARD-Klassik is a joint effort of the editorial departments at the different LRAs, such as SWR, BR, or NDR, as well as the ARD itself, curating and providing genre-related content to their audiences. Since this case study was conducted through interviews with SWR employees and ARD-Klassik producers based at SWR and BR, certain systems and workflows may differ at other LRAs. An overview of the current production network, including SWR and ARD-Klassik, is shown in Fig. 3.

### 4.2.1 Analysis of Workflow

The process analysis begins with the ingest system at the LRA, in this case, SWR Baden-Baden. This system captures everything that the LRA broadcasts. From the ingest system, the LRA editors receive the live-streamed video and audio, which in our case study is the recording of a classical concert. The LRA has an internal wiki used by the editors to store additional information, such as the correct spelling of a conductor's name. Each LRA has its own editing team responsible for preparing the recording for upload to the LRA-specific archive and to the MAM, which serves as a connection between the various LRAs within the ARD. In addition, information about the content and production metadata is sent to the ARD-Klassik editors (refer to the 'ARD Editorial' section in Fig. 3).

The in-house developed middleware service at SWR, AeXpresso, serves as a connection point between the ARD-Klassik editors at SWR and the content delivery network (CDN) as well as the metadata database ARD Core, which is the data and metadata backbone of the ARD Mediathek. AeXpresso is specifically used by SWR to transcode media assets and standardize metadata according to ARD standards.

Within the ARD-Klassik network infrastructure, parts of the concert are re-edited for publishing, submitted to be transcoded and then uploaded to the ARD Mediathek. In the case of an SWR production, this process involves using aeXpresso to open the media files, forwarding them to the transcoding service, and uploading all available metadata to the ARD Core.

The media files themselves are then stored on the CDN system, while the corresponding metadata are externally stored in the ARD Core, at the content-providing LRA and periodically sent to the CMS used by the editors to design the homepage.

At this point the ARD workflow has ended, and the user can access the content through the ARD Mediathek. They can choose the content they want to consume based on e.g. a title, which is stored in the CMS, and then watch the corresponding content, which they receive from the CDN, alongside the necessary metadata for playback, which they receive from the ARD Core through a page- and asset-gateway.

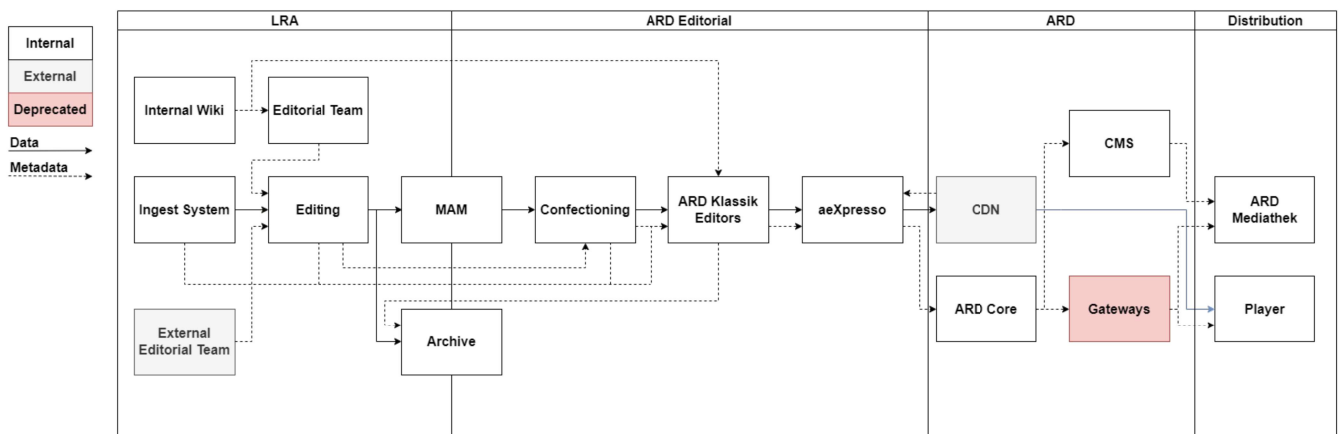

Fig. 3: A simplified representation of the workflow behind an on-demand ARD-Klassik production.

## 4.3 Core Challenges for a Unified Advanced Metadata Implementation

Analyzing the processes, workflows, and infrastructures described above, multiple challenges were identified that needed to be addressed to enable an efficient and holistic approach to metadata management. Although the current workflow facilitates the creation, transportation, and storage of the required metadata, it includes steps and systems that create unnecessary redundancies and increased effort. By addressing these challenges, redundancies can be reduced, and less effort will be needed to create, store, and transport metadata. Furthermore, this will establish a workflow capable of handling the amount of metadata necessary for concepts such as MGA.

The challenges identified in the workflow can be divided into four categories: lack of automation, lack of interoperability, outdated formats and systems, and missing standards. The problems arising from these challenges are categorized based on the metadata assessments described described by Zeng and Qin [5] and Bruce and Hillmann [39].

### 4.3.1 Lack of Automation

There are three available options for extracting metadata from content: manual, automatic, or a combination of both. While manual metadata generation can enhance metadata quality [40], it is also time-consuming and consequently costly [5]. Conversely, automatic metadata extraction facilitates the generation of a larger quantity of metadata but may encounter quality issues. The combination of manual and automatic methods has the potential to enhance both productivity and accuracy, or quantity and quality simultaneously [5]. Presently, a substantial amount of metadata is manually generated, leading to resource challenges and potential drawbacks such as consistency and completeness issues, which will be addressed in the following subsections.

### 4.3.2 Consistency

Consistency challenges arise when the exact same data is recorded differently at different times e.g. in the dc:format an element may be described as "text", "text/plain" or "plain" depending on the element even though all of them share the same format [5] and should have an identical description. According to Zeng and Quin [5] this is most likely due to an absence of standards or recommended guidelines.

A key consistency challenge for ARD-Klassik is the format of the timecodes for different movements of a classical piece of music, such as a symphony or song cycle. The files containing these timecodes do not follow any guidelines or best practices. For example, a timecode might be split into four segments with a reference starting time of 10:00:00:00 (10:17:04:00) or consist of three segments with a reference starting time of 00:00:00 (00:17:04). Furthermore, the files in which this information is stored lack a standardized format, allowing for various file types, such as CSV, TXT, XLSX, or DOCX. These inconsistencies in metadata then need to be manually reformatted and entered into aeXpresso, leading to increased manual processing and, consequently, a greater demand on human resources.

### 4.3.3 Completeness

Completeness, as defined by Zeng and Qin [5], is a measure used to determine the size and distribution of metadata elements in a completed record. This measurement includes the range of descriptive detail in each individual record, and the overall distribution of various levels of records in a metadata database [5].

An example of a completeness problem is the varying level of detail in the timecode files. Some files contain additional metadata, such as the orchestra's name and the conductor's name, while others only include the timecode and name for each movement. Even the level of detail for the names of the movements may vary from one movement to another.

As Bruce and Hillmann [39] describes completeness goals:
- the element set used should describe the target objects as completely as economically feasible [39][p.5]
- the element set should be applied to the target object population as completely as possible [39][p.5]

The given example illustrates that the elements describing the different target objects may vary in their level of detail for each individual object, contrasting with goal 1. Simultaneously, it may also vary from object to object which element is provided, contrasting with goal 2. This leads to a situation, as described by Bruce and Hillmann [39], where a particular element set cannot be reliably used across the entire collection.

### 4.3.4 Interoperability

Interoperability describes the ability of different systems to exchange data. These systems may employ different hardware and software platforms, data structures, and interfaces, but they should aim to exchange the data with minimal loss of content and functionality [41]. As Zeng and Qin [5] also highlight:

> It is becoming generally accepted in the information community that interoperability is one of the most important principles of metadata implementation [5, p.349].

If two given systems in a process cannot communicate data in its entirety, data silos are created, which hinder metadata retrieval and discoverability while simultaneously increasing the resources necessary for the process. 'Data silo' is a term for a segregated group of data stored in different applications [42]. This leads to each user having only limited visibility of the entire data stored across the enterprise. Furthermore, these data silos may have their own data models, vocabularies, formats, and best practices [5]. This situation not only causes redundant processes but can also lead to restricted information sharing, collaboration, and poor decision-making [42].

Since the two metadata databases (ARD Core and the archive) are not connected, discoverability may be affected due to the lack of a complete metadata description in either database. A user might search for certain metadata descrip-tions in the archive, while they are only available in the ARD Core, and vice versa.

### 4.3.5 Resources

A lack of interoperability can also lead to an increase in manual processing. For example, in the ARD workflow, changes to the conductor's name need to be updated in both the archive and the ARD Core. Since the systems cannot exchange information, each database must be updated separately. Furthermore, as described in section 4.3.4, if a journalist is searching for a specific piece of information, it may be necessary to search through all databases (such as the archive and the ARD Core) separately until they find the one containing the metadata they need.

### 4.3.6 Accuracy

If one of the systems is not updated in the process described in section 4.3.5, this leads to inconsistencies and inaccuracies. This would then pose challenges to correctness, as there would be two different metadata descriptions for the same resource.

### 4.3.7 Outdated Formats and Systems

The main challenge concerning outdated formats is the limitation of embedded metadata. For example, if a broadcasting company uses BWF as their format to transport media and metadata, they are limited by the BWF specification (cf. 3.2.1).

The embedded metadata may also be limited by the implementation of the specification used by the software interacting with the files. For instance, even if BWF supports 256 ASCII characters for the description metadata, a software might support more, fewer, or the same number of characters for description metadata in a BWF file. As Lacinak [43] describes, the use of embedded metadata can be severely hindered by the software used to interact not only with the embedded metadata but also with the file in general. Software might ignore metadata embedded in a media file, or, in the worst case, might even delete it entirely.

Based on this information, any format used should be examined not only based on the criterion of meeting the broadcasters' standards according to its specification but also in terms of how well this specification is supported by the systems and software used throughout the process

### 4.3.8 Manual Processing

If a specific system required for the given process simultaneously limits the type or the amount of metadata that can be embedded in the media file, the unsupported metadata must be transported separately to the next system. This may lead to additional manual processing, as the metadata has to be exported, potentially reformatted to fit the criteria of the receiving system, and then imported again

### 4.4 Summary

In the previous sections, several challenges affecting current and future workflows have been identified. By tackling these challenges, broadcasting companies, such as the ARD and the LRA, as well as employees and users, would benefit.

For broadcasting companies and employees, the main benefit would be a reduction in manual processing, specifically in tasks that editors may not enjoy. Connecting systems could eliminate the need for editors to export, standardize, and later import metadata. Editors and journalists would also be able to access all the information they need for their work from a single source, eliminating the need to search through multiple systems. Since the connection between the systems would be standardized, the entire dataset would achieve a certain level of standardization, consistency, and completeness. This would allow prototypes to reliably use specific metadata throughout the entire content catalog.

For users, the benefits include the potential that employees use the resources saved from the reduced necessary manual labor to either create more metadata for personalized content or to produce more content with the current metadata standard. For example, the time currently devoted to standardizing timecode files could instead be used to enrich the timecodes with more information, which could then form the basis for an interactive concert experience.

## 5. Next Steps and Future Perspectives

### 5.1 Metadata and ADM

Although ADM already allows for specific metadata for individual tracks comprising a piece of content, it is currently not possible to further describe segments within these tracks. The next step, as illustrated in the prototypes described in section 4.1, is to enhance the amount of metadata describing parts of a piece of content. This progression involves moving from older metadata standards that describe the entire content, to current standards like ADM, which detail the different tracks comprising the content, and then to a metadata standard that enables the description of segments within these tracks. For example, this would allow for describing specific clips within a track inside the content.

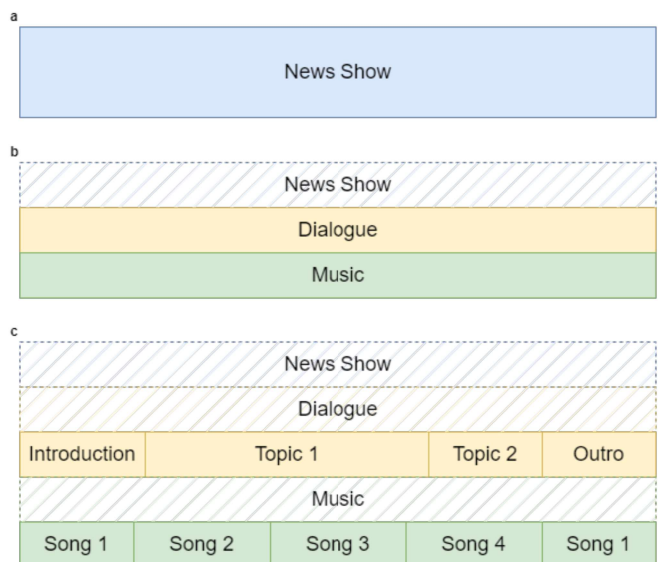

Fig. 4. An overview of how metadata standards are evolving based on a news show example. Every figure showcases the metadata that can be embedded in various metadata standards. (**a**) In outdated formats only metadata about the entirety of the show (e.g. title) can be embedded. (**b**) Apart from general metadata about the news show, current metadata standards such as ADM allow metadata about the individual components i.e. tracks of the news show to be embedded in the file. Examples could be the language of the dialogue. (**c**) Future metadata models further allow the embedding of metadata about parts of the components i.e. clips. An example from a news show could be that for each news segment location metadata is generated and embedded.

An example can be illustrated using a news program: Instead of only storing metadata about the entire program (e.g., the title, a description, its length, etc.), as shown in Fig. 4a, current formats such as BW64 enable the separate storage of information about the track-based components of this news show (Dialogue, Music, etc.), as depicted in Fig. 4b.

A future metadata description may additionally include descriptions of segments within these tracks (e.g., the dialogue consists of an introduction, different topics, an outro, etc.), as depicted in Fig. 4c. It is important to note that the current metadata models, as well as future models, not only store the new metadata but also incorporate metadata from the previous models. For example, the current metadata standard includes fields to store information about the entire piece of content, not just its tracks.

### 5.2 Potential of Artificial Intelligence

Using AI, the extraction of new information about content and thus metadata is possible. Especially in a broadcasting context, the authenticity of content is an important factor. Projects such as Digger by Fraunhofer [44] showcase how AI can be used to analyze the authenticity of content, which could then be stored as metadata embedded in the file for journalists to use in their research.

Voitto from the broadcaster Yle demonstrates the possibility of creating automated, personalized content recommendations based on the metadata stored about the broadcaster's content library, in combination with user preferences [45]. It is a smart assistant that recommends news stories that might be interesting to a user, and based on whether they liked this story or not, it will adapt future recommendations. Furthermore, users can give direct input on what topics they are interested in to influence their recommendations.

Another potential application of AI for the future, as described by Bazán-Gil [46], is in the field of accessibility. An AI could automatically create subtitles and translations, making the content and its information accessible to groups such as the hearing impaired, who might currently be unable to consume the given content. These transcripts could then be used to identify protagonists and keywords for tagging media, which would improve content discovery and visibility, attracting a larger user base [46].

The International Broadcasting Convention (IBC) has also published an article highlighting various use cases of AI in broadcasting, such as automated dubbing, file management, and metadata extraction [47]. Metadata extraction, in particular, enables the generation and utilization of more metadata, as described in the context of Tailored Media in Chapter 4.1. This demonstrates the possibilities of making media archives easier to build and search.

### 5.3 Perspective

Through AI, it is possible to create more metadata with fewer resources. However, the possibilities of AI, and especially metadata, extend beyond just improving discoverability and accessibility. Based on AI-generated transcripts of any given content, chapters, segments, and events can be detected and stored. During playback, these can be used to allow users to, for example, jump to specific moments of a soccer match when a goal was scored. It could also be used in news broadcasting to create a personalized news show based on the interests of the viewer or listener. Segments not matching the user's interests are skipped, while only content relevant to the user is played. The user might even have the option to adapt these segments in length, such as adding or extending sports segments for more detailed coverage.

Further interactivity could include automatic summarization. If the algorithm is aware of all the events in, for example, a soccer match, it could automatically compile a highlight reel of these events for viewers who started watching later. The combination of various segments could also be used to create non-linear stories, allowing users to decide how the story should continue, i.e., which segment should be played next.

For all of these concepts, more metadata is required than is currently generated. While AI-based tools can support the extraction of metadata, the existing systems behind broadcasting also need to adapt to the increasing amount of metadata necessary for adaptive, interactive, and immersive content. Systems need to be updated to support modern formats such as BW64, enable the embedding of more metadata through formats like ADM, and communicate with other systems used in the content creation process. These steps would lead to the generation of more metadata in less time, enabling more personalized content.

By shifting the mindset at broadcasting companies from 'collect as little metadata as is necessary' to 'collect as much metadata as possible,' especially through the use of AI, prototypes based on these or other concepts could be implemented more cheaply and quickly. A process of generating content with the necessary metadata could be entirely bypassed, with the additional benefit that if the prototype is later released as a product, it already has a library of content with the required metadata

**Remark:** In order to enhance the linguistic quality, this text was revised with the assistance of the language models ChatGPT-3.5 and ChatGPT-4. Care was taken to ensure that the meaning of the original formulations was preserved. One section of this text was written based on prompts provided to ChatGPT. This section is explicitly quoted with the exact wording of the prompts, also respecting the use of 'metadata' as a singular expression. The content of the output was verified for accuracy and corrected by the authors when necessary.

[1] R. Gartner, Metadata. Cham: Springer International Publishing, 2016. doi: 10.1007/978-3-319-40893-4.
[2] R. L. Ackoff, "From Data to Wisdom", Journal of Applied Systems Analysis, vol. 16, pp. 3–9, 1989.
[3] S. Hirschmeier, R. Tilly, and V. Beule, "Digital Transformation of Radio Broadcasting: An Exploratory Analysis of Challenges and Solutions for New Digital Radio Services", Jan. 2019. doi: 10.24251/HICSS.2019.602.
[4] R. Studer, V. R. Benjamins, and D. Fensel, "Knowledge engineering: Principles and methods", Data & Knowledge Engineering, vol. 25, no. 1, pp. 161–197, Mar. 1998, doi: 10.1016/S0169-023X(97)00056-6.
[5] M. L. Zeng and J. Qin, Metadata, 2nd ed. Chicago: ALA Neal-Schuman, 2016.
[6] A. Vargas Arcila, S. Baldassarri, and J. Arciniegas, "Study and Comparison of Metadata Schemas for the Description of Multimedia Resources," Apr. 2016, pp. 59–73. doi: 10.1007/978-3-319-38907-3_6.
[7] H.-G. Kim, N. Moreau, and T. Sikora, MPEG-7 audio and beyond: Audio content indexing and retrieval. John Wiley & Sons, 2006.
[8] *EBU Core Metadata Set (EBUCore)*, EBU-TECH 3293, European Broadcasting Union, Geneva, Switzerland, April 2020.
[9] M. Keltsch, "BMF – Metadata Exchange Format of the German Public Broadcasters", EBU Headquarters, 2019.
[10] M. Dickreiter, V. Dittel, W. Hoeg, und M. Wöhr, Handbuch der Tonstudiotechnik, 9., Aktualisierte und Erweiterte Edition. Berlin ; Boston: De Gruyter Saur, 2023.
[11] D. Austerberry, Digital Asset Management. CRC Press, 2012.
[12] *Specification of the Broadcast Wave Format (BWF)*, EBU-TECH 3285, European Broadcasting Union, Geneva, Switzerland, May 2011.
[13] *File format for the exchange of audio programme materials with metadata on information technology media*, Rec. ITU-R BS.1352-4, International Telecommunications Union, Geneva, Switzerland, May 2023.
[14] *RF64: An Extended File Format for Audio Data*, EBU-TECH 3306, European Broadcasting Union, Geneva, Switzerland, June 2018.
[15] "IBM Documentation". [Online]. Available: https://www.ibm.com/docs/en/zos/2.3.0?topic=u6bas-what-is-64-bit-address-space
[16] *Long-form file format for the international exchange of audio programme materials with metadata*, Rec. ITU-R BS.2088-1, International Telecommunications Union, Geneva, Switzerland, Oct. 2019.
[17] *Advanced sound system for programme production*, Rec. ITU-R BS.2051-3, International Telecommunications Union, Geneva, Switzerland, May 2022.
[18] *Material Exchange Format (MXF) - File Format Specification*, ST 377-1:2019, Jan. 2020.
[19] For Television - Material Exchange Format (MXF) - Operational pattern 1A (Single Item, Single Package), ST 378:2004, Sep. 2004.
[20] *Parameters for UHD/HDR Programmes Exchanged as Files*, R 154 v2.0, June 2023.
[21] M. Bolt, "Practical implementation of new open standards for Next Generation Audio production and interchange", European Broadcasting Union, Geneva, Switzerland, 1609-1469, Dec. 2021.
[22] European Broadcasting Union, "Technology Fact Sheet: Next Generation Audio (NGA)", European Broadcasting Union, Geneva, Switzerland, Nov. 2019. [Online]. Available: https://tech.ebu.ch/docs/factsheets/ebu_tech_fs_nga.pdf
[23] *ST 2127-1:2022 - SMPTE Standard - Mapping Metadata-Guided Audio (MGA) signals into the MXF Constrained Generic Container*, ST 2127-1:2022, July 2022, doi: 10.5594/SMPTE.ST2127-1.2022.
[24] *ST 2127-10:2022 - SMPTE Standard - Mapping Metadata-Guided Audio (MGA) signals with S-ADM Metadata into the MXF Constrained Generic Container*, ST 2127-10:2022, July 2022, doi: 10.5594/SMPTE.ST2127-10.2022.
[25] "BBC TASTER" [Online]. Available: https://www.bbc.co.uk/taster
[26] "*Derek Tangye - The Cornish Gardener*" [Online]. Available: https://www.bbc.co.uk/programmes/b02x66zs
[27] "Future Content Experiences: The First Steps For Object-Based Broadcasting" [Online]. Available: https://www.bbc.co.uk/rd/blog/2015-03-future-content-experiences-the-first-steps-for-object-based-broadcasting
[28] "ORPHEUS Project." [Online]. Available: https://orpheus-audio.eu/
[29] W. Bleisteiner, N. Bogaards, M. Meier, N. Färber, and C. Baume, "ORPHEUS D2.5: Final Pilot Progress Report", p. 27, 2018. [Online]. Available: https://zenodo.org/records/1257091
[30] J. Roberts, "Writing for Strategic Communication Industries", The Ohio State University, Available: https://ohiostate.pressbooks.pub/stratcommwriting/

[31] "Reaper | Audio Production Without" [Online]. Available: https://www.reaper.fm/

[32] A. Weller, "Podcasts mit variabler Länge", Tonmeistertagung 31, 2021.

[33] "Home - BBC MakerBox". [Online]. Available: https://www.bbc.co.uk/makerbox

[34] "Adaptive/Perceptive Podcasts" [Online]. Available: https://github.com/bbc/adaptivepodcasting

[35] "Wissen im Vorbeifahren" [Online]. Available: https://www.br.de/unternehmen/inhalt/technik/drive-by-bayern-110.html

[36] "Publikumszentrierte News mit KI" [Online]. Available: https://medium.com/br-next/publikumszentrierte-news-mit-ki-84ae932f6304

[37] "TailoredMedia - Tailored and Agile enrIchment and Linking fOR sEmantic Description of multiMedia" [Online]. Available: https://www.joanneum.at/digital/referenzprojekte/tailoredmedia

[38] "ARD Klassik - Videos der Sendung | ARD Mediathek". [Online]. Available: https://www.ard-klassik.de

[39] T. Bruce und D. Hillmann, "The Continuum of Metadata Quality: Defining, Expressing, Exploiting", Jan. 2004.

[40] J. Weinheimer. "How to keep the practice of librarianship relevant in the age of the Internet." Vine 29 (1999): 14-37.

[41] *Understanding Metadata*. NISO 2004:2, 2004.

[42] J. Patel, "Bridging Data Silos Using Big Data Integration", International Journal of Database Management Systems, vol. 11, pp. 01–06, June 2019, doi: 10.5121/ijdms.2019.11301.

[43] C. Lacinak, "ARSC Technical Committee: A Study of Embedded Metadata Support in Audio Recording Software: Summary of Findings and Conclusions", Association for Recorded Sound Collections, Eugene, OR, USA, Oct. 2011. Available: https://www.arsc-audio.org/pdf/ARSC_TC_MD_Study.pdf

[44] "Digger - Fraunhofer IDMT", Fraunhofer-Institut für Digitale Medientechnologie IDMT. [Online]. Available: https://www.idmt.fraunhofer.de/de/institute/projects-products/projects/digger.html

[45] "Yle NewsWatch's smart Voitto assistant shows recommendations directly on the lock screen" [Online]. Available: https://yle.fi/aihe/artikkeli/2018/10/12/the-first-of-its-kind-in-the-world-yle-newswatchs-smart-voitto-assistant-shows

[46] V. Bazán-Gil, "Artificial intelligence applications in media archives", Profesional de la información / Information Professional, vol. 32, no. 5, Sep. 2023.

[47] "AI in Broadcast: Accelerated Adoption" [Online]. Available: https://www.ibc.org/artificial-intelligence-in-broadcasting/1096.article